\documentclass[lineno]{jfm}

\usepackage{graphicx}
\usepackage{newtxtext}
\usepackage{newtxmath}
\usepackage{natbib}
\usepackage{hyperref}
\usepackage{caption}
\hypersetup{
    colorlinks = true,
    urlcolor   = blue,
    citecolor  = black,
}

\newcommand{\RomanNumeralCaps}[1]

\newcommand{\ddz}{\frac{\mathrm{d}}{\mathrm{d}z}}

\newcommand{\cor}[1]{{#1}}

\shorttitle{Derivation of the GM/R diffusion tensor from QG dynamics}
\shortauthor{J. Meunier, B. Miquel, B. Gallet}

\title{A direct derivation of the Gent-McWilliams/Redi diffusion tensor from quasi-geostrophic dynamics}

\author{Julie Meunier \aff{1},
  Benjamin Miquel \aff{2},
   Basile Gallet\aff{1}
  \corresp{\email{basile.gallet@cea.fr}}}

\affiliation{\aff{1}{Universit\'e Paris-Saclay, CNRS, CEA, Service de Physique de l'Etat Condens\'e, 91191 Gif-sur-Yvette, France.}
\aff{2}{Univ Lyon, CNRS, Ecole Centrale de Lyon, INSA Lyon, Universit\'e Claude Bernard Lyon 1, LMFA, UMR5509, 69130, Ecully, France.}}

\begin{document}
\maketitle


\begin{abstract}
The transport induced by ocean mesoscale eddies remains unresolved in most state-of-the-art climate models and needs to be parameterized instead. The natural scale separation between the forcing and the emergent turbulent flow calls for a diffusive parameterization, where the eddy-induced fluxes are related to the large-scale gradients by a diffusion tensor. The standard parameterization scheme in climate modeling consists in adopting the Gent-McWilliams/Redi (GM/R) form for the diffusion tensor, \cor{initially put forward based on physical intuition and educated guesses before being put on firm analytical footing using thickness-weighted average (TWA).}
In the present contribution we provide a direct derivation of this diffusion tensor from the quasi-geostrophic (QG) dynamics of a horizontally homogeneous three-dimensional patch of ocean hosting a large-scale vertically-sheared zonal flow on the $\beta$ plane. 
\cor{While less general than the TWA approach, the present QG framework leads to rigorous constraints on the diffusion tensor. First, there is no diapycnal diffusivity arising in the QG GM/R tensor for low viscosity and small-scale diffusivities. The diffusion tensor then involves only two vertically dependent coefficients, namely the GM transport coefficient $K_{GM}(z)$ and the Redi diffusivity $K_R(z)$. Secondly, as already identified by previous authors the vertical structures of the two coefficients are related by the so-called Taylor-Bretherton relation. Finally,} while the two coefficients generically differ in the interior of the water column, we show that they are equal to one another near the surface and near the bottom of the domain for low-enough dissipative coefficients. 
We illustrate these findings by numerically simulating the QG dynamics of a horizontally homogeneous patch of ocean \cor{hosting a vertically sheared zonal current resembling} the Antarctic Circumpolar Current.
\end{abstract}

\begin{keywords}
Quasi-geostrophic flows, Geostrophic turbulence, Ocean processes
\end{keywords}


\section{Introduction}

Oceans and planetary atmospheres host currents or jets in thermal-wind balance with meridional buoyancy gradients. This situation is prone to baroclinic instability, however, and the resulting flows are strongly turbulent. In the ocean this turbulence takes the form of `mesoscale' eddies of size comparable to the Rossby deformation radius, a length scale of the order of 15-20 km in the Southern Ocean. While these vortices are key contributors to heat, salt and carbon transport, they are not resolved in state-of-the-art global climate models, and modelers need to parameterize the turbulent transport instead. 
It was soon realized that this turbulent transport is ill-described by standard horizontal diffusion~\citep{Gent11}. Instead, rapid rotation and strong stratification induce quasi-geostrophic (QG) flows in the ocean interior that transport and mix tracers predominantly along density surfaces. \citet{Redi82} thus proposed to relate the sub-grid fluxes to the large-scale gradients through a diffusion tensor that mixes tracers along mean density surfaces only. While avoiding spurious cross-isopycnal mixing, such a parameterization scheme is unable to describe the transport of buoyancy. \citet{Gent90} subsequently proposed to add an extra eddy-induced advective flux ensuring the transport of buoyancy \citep{Gent95}. Following \citet{Griffies98}, these two contributions are conveniently combined into a `Gent-McWilliams/Redi' (GM/R) diffusion tensor that includes both a symmetric part, Redi's isoneutral diffusion scheme, and an anti-symmetric part  encoding GM's advective contribution. 

\cor{A general derivation of the GM/R diffusion tensor was provided by~\citet{Mcdougall2001}. Using thickness-weighted average (TWA) together with an expansion for small fluctuations, they showed that the (GM/R) diffusion tensor governs the eddy-induced fluxes arising in the evolution equation for the thickness-weighted average of the tracer concentration. The GM flux then corresponds to advection by a `quasi-Stokes' velocity, which is directly related to the Gent-McWilliams transport coefficient $K_{GM}$ and can be readily added to the Eulerian-mean velocity to form the residual velocity~\citep{Andrews76}. More recently, \citet{Young12} relaxed the small-fluctuation assumption and derived an exact TWA formulation of the Boussinesq equations, again framed in terms of a three-dimensional residual velocity that includes the eddy-induced advection. Eddy-forcing of the residual velocity arises through the divergence of Eliassen-Palm vectors, which can be recast in terms of the potential vorticity fluxes using some very general form of the Taylor-Bretherton relation (see also Ref.~\cite{Maddison2013} and the discussion around equation (\ref{TBrelation}) below).}

\cor{While they should guide the implementation of parameterizations into global ocean models, such very general TWA formulations do not provide prescriptions for the magnitude or depth-dependence of the eddy-induced fluxes. 
 Instead, progress can be made regarding the overall magnitude and structure of the eddy-induced transport by leveraging the rapid global rotation and focusing on the simpler QG system. Historically, this approach has led to great insight into the overall magnitude of the transport induced by baroclinic turbulence, based on the study of the two-layer QG model~\citep{Phillips,Salmon,Salmon80,Larichev95,Held96,Arbic,Arbic2004b,Arbic2004a,Thompson06,Thompson07,Chang,Gallet2020,Gallet2021}. In a similar fashion, the goal of the present study is to understand the structure of the diffusion tensor in a simple and strongly idealized situation. We thus consider the eddy-induced meridional and vertical transport arising from the QG dynamics of a three-dimensional horizontally homogeneous vertically sheared zonal flow on the $\beta$ plane, see Figure~\ref{fig:schematic}. Baroclinic instability of the base state rapidly leads to turbulence and we wish to characterize the transport properties of the resulting equilibrated state. Focussing on a QG system allows one to make progress regarding the structure of the diffusion tensor, specifically:
\begin{itemize}
\item TWA coincides with standard Eulerian average at fixed depth $z$ in QG (see appendix~\ref{app:McMc}), which allows for a particularly compact derivation of the GM/R diffusion tensor.
\item In QG there is an inverse cascade of energy and buoyancy variance in the interior, with no anomalous dissipation or mixing. Hence, no diapycnal diffusivity coefficient appears in the Redi tensor in the limit of vanishing viscosity and small-scale diffusivities. Together with the (statistical) zonal invariance, this leads to a GM/R diffusion tensor that involves two vertically dependent transport coefficients only, the GM coefficient $K_{GM}(z)$ and the Redi diffusivity $K_R(z)$. By contrast, TWA can be performed both for rotating and non-rotating stratified turbulence and the theory does not dictate \textit{a priori} whether diapycnal fluxes arise (such diapycnal fluxes certainly arise for non-rotating stratified turbulence, see \citet{Linden1979,Peltier2003,Maffioli2016,Caulfield2021}).
\item The QG expression for potential vorticity leads to the Taylor-Bretherton relation between the profiles of $K_{GM}$ and $K_R$ in the interior of the domain.
\item Additionally, the QG boundary conditions at top and bottom readily provide constraints on the top and bottom values of $K_{GM}$ and $K_R$. Namely, these two coefficients are equal at top and bottom for low bottom drag.
\end{itemize}}

\cor{We introduce the theoretical setup in section~\ref{sec:setup}. We highlight the main conservation relations in section~\ref{sec:Material} from which we derive the GM/R diffusion tensor in section~\ref{sec:Arbitrary}. In section~\ref{sec:Constraints} we identify constraints relating the GM coefficient and the Redi diffusivity. We illustrate these constraints using direct numerical simulation in section~\ref{sec:numerics}, before concluding in section~\ref{sec:Conclusion}. In appendix \ref{app:Diff} we show that the contributions from the small-scale diffusive terms are negligible in the QG regime. In appendix~\ref{app:McMc} we make the connection between the present QG approach and the more general TWA approach of \citet{Mcdougall2001}. Finally, the details of the numerical procedure are provided in appendix~\ref{app:DNS}.}


\section{Quasi-geostrophic dynamics of an idealized 3D patch of ocean\label{sec:setup}}

\begin{figure}
    \centerline{\includegraphics[width=14 cm]{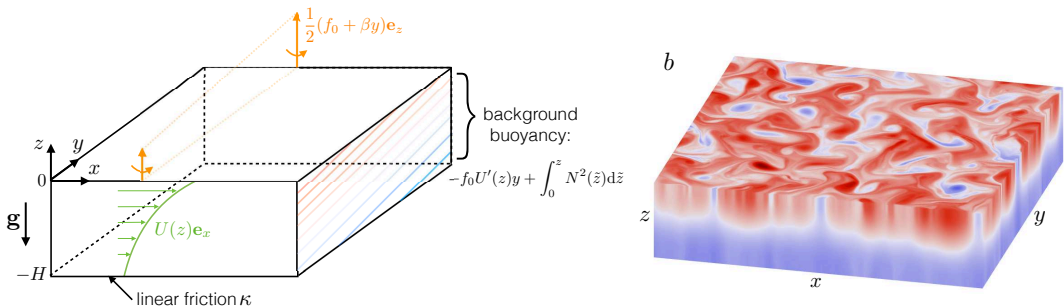} }
   \caption{\textbf{Left:} An idealized patch of ocean. A layer of fluid is subject to global rotation at a rate that varies linearly with the meridional coordinate $y$. The fluid is density-stratified with a profile $N(z)$ for the buoyancy frequency. The background zonal shear flow has a profile $U(z)$. The flow coexists with a background meridional buoyancy gradient as a result of thermal-wind balance. Bottom friction damps the fluctuating kinetic energy. \textbf{Right:} Snapshot of the departure buoyancy field $b$ in the equilibrated state of the numerical simulation.\label{fig:schematic}}
\end{figure}

We consider the idealized patch of ocean represented in figure~\ref{fig:schematic}. Water occupies a volume $(x,y,z)\in [0,L]^2 \times [-H,0]$ with a stress-free boundary at the surface $z=0$ and a linear-friction boundary condition at $z=-H$. The fluid layer is subject to global rotation around the vertical axis with a local Coriolis parameter $f_0+\beta y$, where $y$ denotes the meridional (North-South) coordinate. Additionally, the fluid layer is density-stratified with an arbitrary buoyancy frequency profile $N(z)$, and we restrict attention to a single stratifying agent. We focus on the rapidly rotating strongly stratified regime for which the fluid motion is governed by quasi-geostrophy~\citep{Venaille11,Salmonbook,Vallisbook}. In that limit the velocity field ${\bf u}=(u,v,w)$ consists of a leading-order horizontal geostrophic flow $(u,v)=(-P_y,P_x)$, where the generalized pressure field $P$ is defined as the opposite of the streamfunction, together with subdominant vertical velocity $w$. The buoyancy field is given by $B=f_0 P_z$ as a result of hydrostatic balance.

The base flow consists of an arbitrary zonal velocity profile $U(z)=-P_y$. Differentiating with respect to $z$ indicates that the zonal flow is in thermal wind balance with a $z$-dependent meridional buoyancy gradient, $\partial_y B=-f_0 U’(z)$. We consider the evolution of arbitrary departures from this base state. We denote as $p(x,y,z,t)$ the departure from the base pressure field, with $u=-p_y$ the departure zonal velocity, $v=p_x$ the departure meridional velocity and $b=f_0 \, p_z$ the departure buoyancy. In the following we adopt dimensionless variables, with time expressed in units of $|f_0|^{-1}$ and lengths in units of $H$. 
For brevity we use the same symbols for the dimensionless variables. 
The QG limit is obtained for small isopycnal slope of the base state, or equivalently in the small Rossby number limit for ${\cal O} (1)$ stratification. Denoting as $\epsilon$ the typical magnitude of the isopycnal slope, the QG system can be derived through the following scalings: 
\begin{eqnarray}
& & N^2 \sim 1 \, , \qquad  \beta \sim \epsilon \, , \qquad \partial_x, \partial_y, \partial_z \sim 1 \, , \qquad \partial_t \sim \epsilon \, , \label{scalings}\\
\nonumber & & U \sim \epsilon \, , \qquad (u,v) \sim \epsilon \, , \qquad w \sim \epsilon^2 \, , \qquad b \sim \epsilon \, .
\end{eqnarray}
A standard asymptotic expansion of the equations of motion leads to the QG system~\citep{Pedloskybook,Salmonbook,Vallisbook}, as recalled in \citet{Gallet2022} for the specific notations and scalings considered here.  The evolution then reduces to a conservation equation for the quasi-geostrophic potential vorticity (QGPV). The dimensionless QGPV departure $q$ is related to the departure pressure $p$ through:
\begin{equation}
q = \Delta_\perp p + \partial_z \left( \frac{p_z}{N^2(z)} \right) \, , \label{eq:defq}
\end{equation}
where $\Delta_\perp=\partial_{xx}+\partial_{yy}$.
The (dimensionless) QGPV conservation equation reads:
\begin{equation}
\partial_t q + U(z) \, q_x + J(p,q)  = [-{\beta}+{\cal S}'(z)] p_x + {\cal D}_q \, , \label{eq:consq}
\end{equation}
where the Jacobian is $J(g,h)=g_x h_y - g_y h_x$ and we denote the isopycnal slope of the base state as ${\cal S}(z)=U'/N^2$. 
The first term on the right-hand side of (\ref{eq:consq}) corresponds to the distortion of the background meridional PV gradient by the meridional flow. The second term, ${\cal D}_q$, is the contribution from the viscosity and buoyancy diffusivity, which damp the small-scale vorticity and buoyancy fluctuations. The latter are explicited in appendix \ref{app:Diff}.


At the same level of approximation, the evolution equation for the (dimensionless) buoyancy departure $b=p_z$ reads:
\begin{equation}
\partial_t b + U(z) \, b_x + J(p,b) =  U'(z) p_x - w N^2(z)  + {\cal D}_b \, , \label{eq:consb}
\end{equation}
where the diffusive term ${\cal D}_b$ is provided in appendix \ref{app:Diff}. The first two terms on the right-hand side are the sources of buoyancy fluctuations in the system: they correspond to the distortion by the turbulent flow of the background meridional and vertical buoyancy gradients, respectively.
At the surface, where $w=0$, this equation reduces to:
\begin{equation}
\partial_t b|_0 + U(0)  \, b_{x}|_0 + J(p|_0, b|_0) = U'(0) p_x|_0 + {\cal D}_b|_0 \, , \label{eq:bz1}
\end{equation}
where the subscript $\cdot |_0$ refers to quantities evaluated at $z=0$.
Quantities evaluated just above the bottom Ekman boundary layer are denoted with the subscript $-1^+$. At this depth, the pumping vertical velocity is given by $w|_{-1^+}=\kappa \Delta_{\perp} p_{-1^+}$, where the friction coefficient $\kappa$ can either be related to the vertical viscosity through standard Ekman layer theory over a flat bottom boundary, or specified at the outset as an independent coefficient parameterizing more realistic drag on the ocean floor (see appendix \ref{app:DNS} and \citet{Gallet2022}). The evolution equation for the buoyancy at $z=-1^+$ reads:
\begin{eqnarray}
& & \partial_t b|_{-1^+} + U(-1) \, b_{x}|_{-1^+}  +  J(p|_{-1^+},b|_{-1^+})   \label{eq:bz0}  \\
\nonumber & &= U'(-1) p_x|_{-1^+} - N^2(-1) \kappa \Delta_{\perp} p|_{-1^+} +  {\cal D}_b|_{-1^+} \, .
\end{eqnarray}
One way to integrate the QG system consists in marching in time the QGPV conservation equation (\ref{eq:consq}) together with the top and bottom buoyancy equations (\ref{eq:bz1}) and (\ref{eq:bz0}). To infer the pressure field at each time step, one inverts the relation (\ref{eq:defq}) using $b|_{-1^+}$ and $b|_{0}$ as boundary conditions. A desirable feature of this QG approach is that it is fully compatible with periodic boundary conditions in $x$ and $y$ for the departure fields. We adopt such periodic boundary conditions in the following.

In the bulk of the domain, the buoyancy evolution equation (\ref{eq:consb}) provides a diagnostic relation to infer the subdominant vertical velocity $w$, the latter being crucial to parameterize eddy-induced vertical transport.

\section{Material invariants: buoyancy, QGPV and the cross-invariant\label{sec:Material}}

We denote with an overbar $\overline{\cdot}$ a time average together with a horizontal area average. Our goal is to characterize the transport properties of the flow, more specifically the diffusion tensor connecting the eddy-induced fluxes to the large-scale background gradients of some arbitrary tracer, be it active or passive. One can gain insight into the structure of this tensor by focusing on two specific tracers: buoyancy and QGPV. In this section we thus derive rigorous constraints between the meridional and vertical turbulent fluxes of buoyancy and QGPV: $\overline{vb}(z)$, $\overline{wb}(z)$, $\overline{vq}(z)$ and $\overline{wq}(z)$.

The first constraint stems from the conservation of buoyancy variance. Multiplying the buoyancy evolution equation (\ref{eq:consb}) with $b$ before averaging over time, $x$ and $y$ leads to:
\begin{equation}
N^2 \overline{wb} = U'  \overline{vb} \, , \label{eq:tempb2}
\end{equation}
up to diffusive corrections that vanish in the regime of low viscosity and buoyancy diffusivity. 
A proof that the diffusive contributions indeed vanish is provided in appendix \ref{app:Diff} based on the well-known absence of a forward energy cascade in QG turbulence. 
We recast the equality (\ref{eq:tempb2}) in the form:
\begin{equation}
\overline{wb} =  {\cal S}(z) \overline{vb} \, , \label{isotransport}
\end{equation}
which shows that the mean buoyancy transport is directed along the mean isopycnals.

We derive a second constraint on the fluxes based on the conservation of the cross-invariant $bq$. Multiply the QGPV evolution equation (\ref{eq:consq}) with $b$ and the buoyancy conservation equation (\ref{eq:consb}) with $q$. Summing the resulting equations and averaging over time, $x$ and $y$ leads to:
\begin{equation}
\overline{wq} =  {\cal S}(z) \overline{vq} - \frac{\beta-{\cal S}'(z)}{N^2(z)} \overline{vb} \, , \label{crossrel}
\end{equation}
where the equality holds in the low-diffusivity limit, where, as shown in appendix \ref{app:Diff}, the contributions from viscosity and buoyancy diffusivity vanish.

\section{Arbitrary tracer: Gent-McWilliams/Redi diffusion tensor\label{sec:Arbitrary}}

Define the $z$-dependent eddy diffusivities $K_R(z)$ and $K_{GM}(z)$ as:
\begin{eqnarray}
K_R(z)  =  \frac{\overline{vq}}{-\beta+{\cal S}'(z)} \qquad \text{and} \qquad K_{GM}(z)  =  \frac{\overline{vb}}{U'(z)} \, , \label{defK}
\end{eqnarray}
namely, $K_R$ is the ratio of the meridional PV flux over minus the background meridional PV gradient, while $K_{GM}$ is the ratio of the meridional buoyancy flux over minus the background meridional buoyancy gradient. The notations $K_{GM}$ and $K_R$ will become obvious at the end of the derivation to come.

Now consider a tracer $\tau$ stirred by the 3D flow and subject to horizontally uniform gradients $G_y(z)={\cal O}(\epsilon)$ and $G_z(z)={\cal O}(1)$ in the meridional and vertical directions, respectively. \cor{That is, the total tracer field reads:
\begin{eqnarray}
\int_{-1}^z G_z(\tilde{z}) \, \mathrm{d}\tilde{z} + y \, G_y(z) + \tau(x,y,z,t) \, .
\end{eqnarray}}

Under these conditions and with the scalings (\ref{scalings}) the evolution equation for $\tau$ reads:
\begin{equation}
\partial_t \tau + U(z) \, \tau_x + J(p,\tau) =  - p_x G_y(z) - w G_z(z)  + {\cal D}_\tau \, , \label{eq:constau}
\end{equation}
where ${\cal D}_\tau$ denotes small-scale diffusion. A few remarks are in order regarding the background meridional and vertical gradients: $G_y$ and $G_z$ above should be understood as the lowest-order background gradients that enter QG dynamics. \cor{Naturally, a subdominant vertical gradient $G_z^{(1)}= y \, \partial_z G_y(z)={\cal O}(\epsilon)$} exists to ensure the equality of the cross-derivatives, $\partial_y(G_z+G_z^{(1)})=\partial_z G_y$. However, one can easily check that $G_z^{(1)}$ is subdominant and does not arise in the QG evolution equation (\ref{eq:constau}). Similarly, keeping $G_z^{(1)}$ on the right-hand side of equation~(\ref{GMRedi}) would lead to negligible corrections to the fluxes, of higher order in $\epsilon$. 


The meridional and vertical fluxes of $\tau$ are related to the background meridional and vertical gradients $G_y$ and $G_z$ through a diffusion tensor:
\begin{eqnarray}
\left( \begin{matrix}
\overline{v\tau} \\
\overline{w\tau}
\end{matrix}  \right) =
\left[\begin{matrix}
A_1(z) &  A_2(z) \\
A_3(z)& A_4(z) \\
\end{matrix} \right]
\left( \begin{matrix}
G_y \\
G_z
\end{matrix}  \right)  \, ,\label{tensordef}
\end{eqnarray}
where the $A_i(z)$ are unknown $z$-dependent coefficients at this stage. Apply relation (\ref{tensordef}) to the tracers $b$ and $q$, the associated background gradients being readily inferred from the right-hand-side terms of equations (\ref{eq:consq}) and (\ref{eq:consb}): $G_y=-U'(z)$ and $G_z=N^2$ for $\tau=b$, and $G_y=\beta - {\cal S}'(z)$, $G_z=0$ for $\tau=q$. We obtain the following fluxes:
\begin{eqnarray}
\left( \begin{matrix}
\overline{vb} \\
\overline{wb}
\end{matrix}  \right) =
\left( \begin{matrix}
-A_1 U'(z) + A_2 N^2\\
-A_3 U'(z) + A_4 N^2
\end{matrix}  \right)  \, ; 
\left( \begin{matrix}
\overline{vq} \\
\overline{wq}
\end{matrix}  \right) =
\left( \begin{matrix}
A_1 [\beta - {\cal S}'(z)] \\
A_3 [\beta - {\cal S}'(z)]
\end{matrix}  \right) .
\end{eqnarray}
There are four constraints on these four fluxes, which allow us to express the four coefficients $A_i$ in terms of $K_{GM}(z)$ and $K_{R}(z)$: the first two constraints are simply the definitions (\ref{defK}) of $K_{GM}$ and $K_{R}$, the third constraint is (\ref{isotransport}), namely the fact that the mean transport of $b$ is along the mean isopycnals, and the fourth constraint is the cross-invariant relation (\ref{crossrel}). After a straightforward calculation of the coefficients $A_i$ the diffusion tensor connecting the fluxes to the background gradients becomes:
\begin{eqnarray}
\left( \begin{matrix}
\overline{v\tau} \\
\overline{w\tau}
\end{matrix}  \right) =
\left[\begin{matrix}
-K_R &  (K_{GM}-K_R) {\cal S}\\
-   (K_{GM}+K_R){\cal S}& - K_R {\cal S}^2\\
\end{matrix} \right]
\left( \begin{matrix}
G_y \\
G_z
\end{matrix}  \right)  \, .\label{GMRedi}
\end{eqnarray}
This form for the diffusion tensor corresponds to the Gent-McWilliams/Redi parameterization (GM/R in the following), where $K_{GM}(z)$ denotes the $z$-dependent Gent-McWilliams coefficient and $K_R(z)$ denotes the Redi diffusivity. The former represents the skew flux associated with adiabatic transport by the eddying flow~\citep{Griffies98}. \cor{Using the definition (\ref{defK}) of $K_{GM}$, we check in appendix~\ref{app:McMc} that the GM part of the tensor (\ref{GMRedi}) corresponds to the QG limit of the advective fluxes associated with the more general quasi-Stokes streamfunction introduced by~\cite{Mcdougall2001}. The Redi part of the tensor represents mixing along the neutral direction in the limit of weak isopycnal slope ${\cal S}(z)$.  As can be inferred from the QGPV conservation equation, the Redi diffusivity $K_R(z)$} also equals the Taylor-Kubo eddy diffusivity coefficient deduced at any height $z$ from the Lagrangian correlation function of the horizontal QG velocity field. Several similar estimates for the PV diffusivity have been compared in channel geometry by~\citet{Abernathey13}. 



\section{Constraints on the GM and Redi coefficients\label{sec:Constraints}}

The derivation above allows us to obtain constraints on the GM and Redi coefficients as defined by (\ref{defK}). First of all, at the upper boundary the buoyancy equation (\ref{eq:bz1}) has exactly the same structure as the QGPV conservation equation (\ref{eq:consq}). Both equations correspond to advection by the base zonal flow and the QG flow at the upper boundary, with a source term that corresponds to the distortion of a background meridional gradient. We conclude that the diffusivities relating the meridional flux to the background meridional gradient are equal for $q$ and $b$ at $z=0$ (and given by the Taylor-Kubo eddy diffusivity coefficient associated with the surface horizontal flow). Using the definitions (\ref{defK}) this leads to the constraint:
\begin{equation}
K_{GM}(0)  =  K_R(0) \, . \label{eqK1}
\end{equation}
The same relation holds near the bottom boundary (just above the Ekman boundary layer) when the drag coefficient is low:
\begin{equation}
K_{GM}(-1^+) = K_R(-1^+) \, . \label{eqK0}
\end{equation}
The equality of $K_{GM}(z)$ and $K_R(z)$ is a common assumption when implementing the GM/R parameterization in global models~\citep{Griffies98}. We put this assumption on firmer analytical footing by showing that it holds near the upper and lower boundaries, although the two coefficients generically differ in the interior of the fluid column.
In the general case, however, we can further relate the vertical dependence of $K_{GM}(z)$ and $K_R(z)$ through the Taylor-Bretherton relation. Multiplying equation (\ref{eq:defq}) with $v=p_x$ before averaging horizontally yields, after a few integrations by parts in the horizontal directions:
\begin{equation}
\overline{vq}=\frac{\mathrm{d}}{\mathrm{d} z} \left( \frac{\overline{vb}}{N^2} \right) \, .\label{QGrel}
\end{equation}
This equation corresponds to the horizontal average of a more general relation initially derived by Bretherton~\citep{Bretherton66} and often referred to as the Taylor-Bretherton relation~\citep{Taylor15,Dritschel08,Young12}. 
Using the definitions (\ref{defK}) we recast (\ref{QGrel}) as:
\begin{equation}
K_R({\cal S}'-\beta)=\frac{\mathrm{d}}{\mathrm{d} z} (K_{GM} \, {\cal S}) \, . \label{TBrelation}
\end{equation}
This equality was previously obtained by several authors (e.g. \citet{Smith09}) and is recalled here for the sake of completeness only.

\section{A numerical example\label{sec:numerics}}

With the goal of illustrating the results above, we turn to the Direct Numerical Simulation (DNS) of an isolated horizontally homogeneous patch of ocean. The setup is chosen to reproduce conditions in the Antarctic Circumpolar Current (ACC), with surface-intensified stratification, shear and turbulence. The base state consists of a (dimensionless) stratification decreasing linearly with depth, from $N^2(0)=400$ at the surface to $N^2(-1)=50$ at the bottom, together with an exponential profile for the background shear flow, $U(z)=Ro \, e^{2z}$ with $Ro=0.03$, and a dimensionless planetary vorticity gradient $\beta=4.0 \times 10^{-5}$. These values for $Ro$ and $\beta$ are ten times smaller than typical values in the ACC: this choice leaves invariant the dissipation-free QG dynamics while ensuring that the numerical simulation is indeed performed in the fully QG regime. In other words, we simulate an idealized horizontally homogeneous and fully QG version of the ACC. We have also kept $f_0>0$, which conveniently leads to $\overline{vb}>0$ while being equivalent to the ACC situation up to an equatorial symmetry. We solve for the fully nonlinear evolution of the departures from the base state inside a domain of dimensionless size $(x,y,z)\in [0;500]^2\times [-1;0]$ using periodic boundary conditions in the horizontal directions and small values for the dissipative coefficients. The numerical procedure is detailed in appendix~\ref{app:DNS}.

\begin{figure}
   \centerline{\includegraphics[width=12.5 cm]{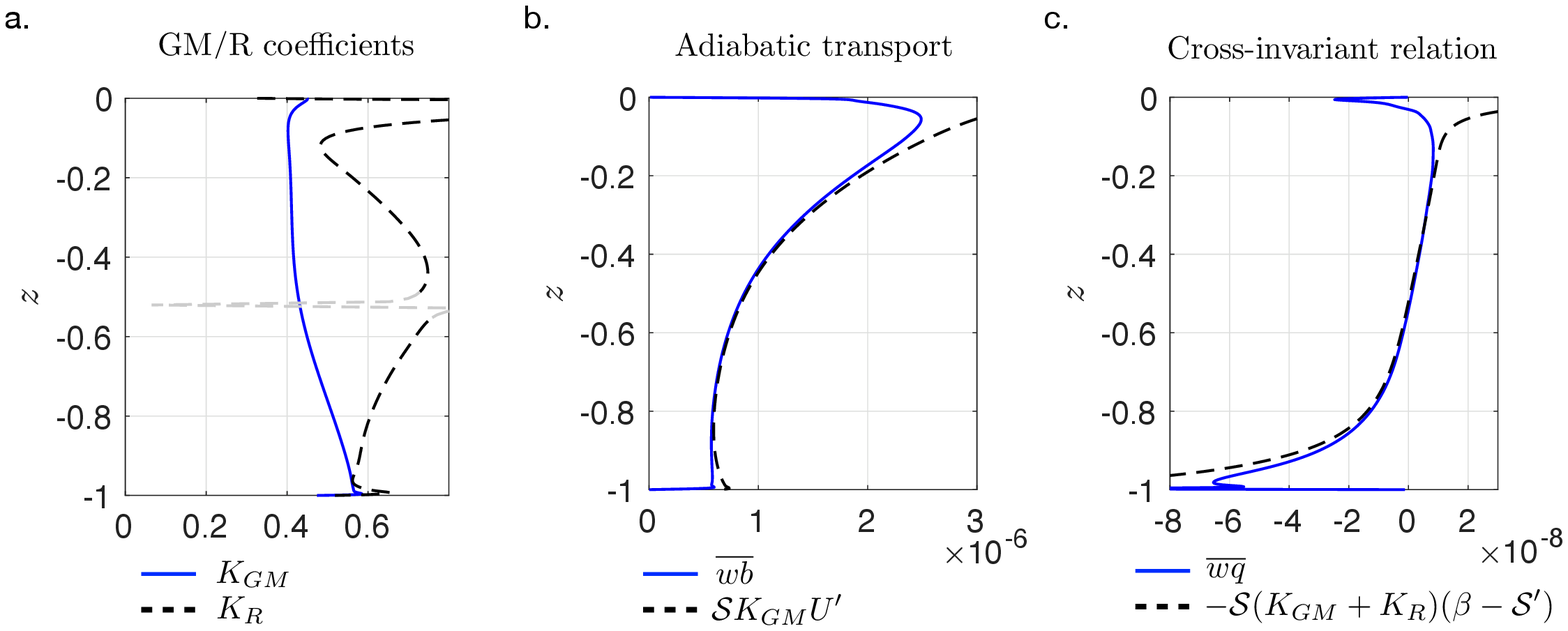} }
   \caption{Time and horizontally averaged profiles from the numerical simulation. \textbf{a.} Transport coefficients as defined in (\ref{defK}) (gray region corresponding to the depth where the meridional QGPV gradient vanishes). \textbf{b.} Vertical buoyancy flux compared to the GM/R prediction, which validates adiabatic transport in the interior. \textbf{c.} Vertical QGPV flux compared to the GM/R prediction, which validates the cross-invariant relation~(\ref{crossrel}). \label{fig:profiles}}
\end{figure}

After some transient the system reaches a statistically steady equilibrated state, illustrated in Figure~\ref{fig:schematic} through a snapshot of the departure buoyancy field $b$. We extract the time and horizontal area averages of the meridional and vertical fluxes of buoyancy and QGPV in this statistically steady state. The corresponding profiles are shown in Figure~\ref{fig:profiles}, the first panel of which provides the diagnosed Gent-McWilliams coefficient and Redi diffusivity. In agreement with results from re-entrant channel simulations~\citep{Abernathey13}, $K_{GM}(z)$ is monotonic in the interior of the domain, whereas $K_R(z)$ exhibits a maximum at mid-depth~\citep{Treguier99,Smith09}. 
In line with the constraints (\ref{eqK1}) and (\ref{eqK0}), the interior profiles of $K_{GM}(z)$ and $K_R(z)$ tend to a common limiting value as we approach the top or bottom boundary. This tendency is disrupted by diffusive boundary-layer effects in the immediate vicinity of the boundaries (more strongly so near the surface). These boundary layers shrink as we lower the diffusivities employed in the numerical simulation.


Having diagnosed $K_{GM}(z)$ and $K_R(z)$ we turn to the vertical fluxes of buoyancy and QGPV, with the goal of comparing the numerical fluxes to the predictions of the GM/R diffusion tensor. Panel \ref{fig:profiles}b shows  that the vertical buoyancy flux $\overline{wb}$ is accurately captured by the GM/R prediction ${\cal S}K_{GM}U'(z)$. Because ${\cal S}K_{GM}U'(z)={\cal S} \overline{vb}$, this validates the fact that buoyancy is transported adiabatically in the interior, in line with equation~(\ref{isotransport}). Panel \ref{fig:profiles}c shows that the vertical QGPV flux $\overline{wq}$ is accurately captured by the GM/R prediction $-{\cal S}(K_{GM}+K_R)(\beta-{\cal S}')$, the latter expression being also equal to the right-hand side of the cross-invariant relation (\ref{crossrel}). The good agreement in panel \ref{fig:profiles}c thus validates the simple and exact cross-invariant relation~(\ref{crossrel}) in the interior of the domain.

\section{Conclusion\label{sec:Conclusion}}

\cor{We have studied the transport properties of the turbulent QG flow arising from the baroclinic instability of a horizontally homogeneous vertically sheared zonal current. 
While less general than the TWA formulation of the Boussinesq equations~\citep{Mcdougall2001,Young12}, the QG limit allows one to make progress on the structure of the diffusion tensor relating the eddy-induced fluxes to the background gradients. Based on the conservation of buoyancy variance and of a cross-invariant involving buoyancy and QGPV, we thus derived a particularly simple GM/R form for the diffusion tensor. First, in the interior of the domain there are no diapycnal fluxes provided the viscosity and small-scale diffusivities are small. The diffusion tensor then involves only two vertically dependent coefficients: the GM transport coefficient $K_{GM}(z)$ and the Redi diffusivity $K_R(z)$. Secondly, based on the definition of QGPV one can relate $K_{GM}$ and $K_R$ through the Taylor-Bretherton relation (\ref{TBrelation}). Finally, based on the QGPV and buoyancy evolution equations one obtains that $K_{GM}$ and $K_R$ are equal to one another at top and bottom. These results provide some support for the common modeling assumption $K_{GM} \simeq K_R$ near the boundaries~\citep{Griffies98}. However, the two coefficients are allowed to depart from one another in the interior of the fluid column and indeed they do in the present numerical simulation (in line with previous studies, see \citet{Abernathey13}). It would be interesting to investigate whether some equivalent of the boundary relation $K_{GM} \simeq K_R$ exists beyond the present idealized QG framework, for a primitive-equation or Boussinesq system. TWA would likely play a central role for such an extension.}

\textbf{Acknowledgements}. We thank G. Hadjerci, R. Ferrari and W.R. Young for insightful discussions. This research is supported by the European Research Council under grant agreement FLAVE 757239. The numerical study was performed using HPC resources from GENCI-CINES and TGCC (grants 2021-A0102A12489, 2022-A0122A12489 and 2023-A0142A12489).

\appendix

\section{Diffusive contributions\label{app:Diff}}

We consider the impact of the standard diffusive terms (viscosity and buoyancy diffusivity) within the framework of QG. We use different coefficients for the diffusivities in the horizontal and vertical directions: $E_{\perp}$ and $E_z$, respectively, for the horizontal and vertical dimensionless viscosities (Ekman numbers), and $E_{b,\perp}$ and $E_{b,z}$, respectively, for the horizontal and vertical dimensionless buoyancy diffusivities. With these notations, the diffusive term in the buoyancy equation (\ref{eq:consb}) reads:
\begin{eqnarray}
{\cal D}_b & = &  E_{b,\perp} \Delta_{\perp} b + E_{b,z} \partial_{zz} b \, ,
\end{eqnarray}
while the diffusive term in the QGPV conservation equation (\ref{eq:consq}) reads:
\begin{equation}
{\cal D}_q  =   E_{\perp} \Delta_{\perp}^2 p + E_z \Delta_{\perp} p_{zz} + \partial_z \left[ \frac{1}{N^2} \left( E_{b,\perp} \Delta_{\perp} b + E_{b,z} \partial_{zz} b \right)  \right]     \, .
\end{equation}
In contrast to standard 3D turbulence, quasi-geostrophic dynamics are characterized by an inverse energy cascade~\citep{Charney71}, together with a forward cascade of buoyancy variance at the boundaries only: in the limit where the various diffusive coefficients $E_i$ are sent to zero simultaneously, there is no `anomalous' energy dissipation and no `anomalous' dissipation of buoyancy variance in the interior~\citep{Lapeyre17}. That is, the limit $E_i \overline{(\Delta_\perp p)^2} \to 0$ holds for any $z$, and the limits $E_i \overline{|\bnabla_\perp b|^2} \to 0$ and $E_i \overline{(b_{z})^2} \to 0$ hold pointwise for $z \neq \{-1; 0\}$. Additionally, any $z$-derivative of these profiles also vanishes  in the vanishing-diffusivity limit for $z \neq \{-1; 0\}$. The diffusive contribution to the right-hand side of the adiabatic-transport relation (\ref{isotransport}) reads:
\begin{eqnarray}
\overline{b{\cal D}_b} & = &  - E_{b,\perp} \overline{|\bnabla_\perp b|^2} - E_{b,z} \overline{(b_{z})^2} + \frac{E_{b,z}}{2} \ddz \overline{b^2} \, .
\end{eqnarray}
The three terms vanish in the vanishing-diffusivity limit for $z \neq \{-1; 0\}$, leading to $\overline{b{\cal D}_b} \to 0$ and relation (\ref{isotransport}). The diffusive contribution to the right-hand side of the cross-invariant relation (\ref{crossrel}) reads $(\overline{b{\cal D}_q}+\overline{q{\cal D}_b})/N^2$, where:
\begin{eqnarray}
\nonumber & & \overline{q{\cal D}_b}  =  \frac{E_{b,\perp}}{2} \ddz \overline{(\Delta_{\perp} p)^2} + \frac{E_{b,z}}{2} \left[ - \frac{\mathrm{d}^3}{\mathrm{d}z^3} \overline{(\bnabla_\perp p)^2} + 3 \ddz \overline{(\bnabla_\perp b)^2}  \right]  \label{eq:qDb} \\
\nonumber & & + E_{b,\perp} \left\{ - \ddz \left[ \frac{\overline{(\bnabla_\perp b)^2}}{N^2}  \right]   + \frac{1}{2N^2} \ddz \overline{(\bnabla_\perp b)^2} \right\} + E_{b,z} \left\{  \left[\frac{\mathrm{d}^2}{\mathrm{d}z^2} \left(\frac{\overline{b^2}}{2}\right) -\overline{(b_z)^2}  \right] \ddz \left( \frac{1}{N^2} \right)  +\frac{1}{2 N^2} \ddz \overline{b_z^2} \right\} \, ,\end{eqnarray}
\begin{eqnarray}
\nonumber \overline{b{\cal D}_q}  & = & \frac{E_\perp}{2} \ddz \overline{(\Delta_\perp p)^2} - \frac{E_z}{2} \ddz \overline{(\bnabla_\perp b)^2} \\
\nonumber & & + \ddz \left[-\frac{E_{b,\perp}}{N^2} \overline{(\bnabla_\perp b)^2}  + \frac{E_{b,z}}{2 N^2} \frac{\mathrm{d}^2}{\mathrm{d}z^2} \overline{b^2} - \frac{E_{b,z}}{N^2} \overline{(b_z)^2}  \right] +  \frac{E_{b,\perp}}{2 N^2} \ddz   \overline{(\bnabla_\perp b)^2} - \frac{E_{b,z}}{2 N^2} \ddz \overline{(b_z)^2} \, .
\end{eqnarray}
The terms on the right-hand side of both expressions vanish in the vanishing-diffusivity limit for $z \neq \{-1; 0\}$, leading to $\overline{q{\cal D}_b} + \overline{b{\cal D}_q} \to 0$. Hence the approximate relation (\ref{crossrel}) for low diffusivities. 



\section{\cor{Connection to TWA and the residual-mean approach}\label{app:McMc}}



\subsection{The evolution equations for the TWA and for the standard fixed-z-averaged tracer concentration are identical in the QG limit}

\cor{For a given tracer $\tau$, \citet{Mcdougall2001} consider the evolution equations for the TWA $\hat{\tau}$ and for the standard fixed-$z$ average $\overline{\tau}$, where the time average is to be understood as an average over a few eddy turnover times. They show that the two evolution equations differ by the divergence of some vector ${\bf E}$, see their equations (53-55). With the QG scalings (\ref{scalings}), however, this additional term is negligible. The horizontal components of ${\bf E}$ are ${\cal O}(\epsilon^3)$, smaller than the meridional flux $\overline{v \tau}={\cal O}(\epsilon^2)$ discussed in the present study. The vertical component of ${\bf E}$ is the time-derivative of some ${\cal O}(\epsilon^2)$ material invariant, averaged over the slow QG eddy turnover timescale $1/\epsilon$. The latter time derivative is thus at least of order $\epsilon^4$, much smaller than the vertical flux $\overline{w \tau}={\cal O}(\epsilon^3)$ discussed in the present study. The vector ${\bf E}$ is thus a higher-order term that is negligible at the level of the QG approximation. In other words, the evolution equations for the TWA and the fixed-$z$ average are identical at the level of the QG approximation.}

\subsection{The coefficient $K_{GM}$ of the present study describes advection by the quasi-Stokes velocity in the QG limit}

\cor{In the general TWA formulation of \citet{Mcdougall2001}, the fluxes arising from the antisymmetric part of the diffusion tensor are expressed in terms of a quasi-Stokes streamfunction $\boldsymbol{\psi}$ as:
\begin{eqnarray}
\left[\begin{matrix}
0 & -\boldsymbol{\psi} \cdot {\bf e}_y \\
 \boldsymbol{\psi} \cdot {\bf e}_y & 0 \\
\end{matrix} \right]
\left( \begin{matrix}
G_y \\
G_z
\end{matrix}  \right)  \, , \label{McMcantisym}
\end{eqnarray}
where we restrict attention to the case of zonally invariant statistics. With the QG scalings (\ref{scalings}) the $y$ component of the quasi-Stokes streamfunction provided in \citet{Mcdougall2001} reduces to $\boldsymbol{\psi}\cdot {\bf e}_y=-{\overline{vb}}/{N^2} + {\cal O}(\epsilon^3)$. 
Using our definition (\ref{defK}) for $K_{GM}$, we can re-express the right-hand side as $- K_{GM} {\cal S} + {\cal O}(\epsilon^3)$. This shows that (\ref{McMcantisym}) is indeed equivalent to the GM part of the tensor (\ref{GMRedi}) of the present study at the QG level of approximation.}


\section{Direct numerical simulation\label{app:DNS}}

The numerical simulations are performed using an intermediate set of equations between the Boussinesq equations and the QG system. Indeed,
on the one hand the QG system -- equation (\ref{eq:consq}) with the boundary conditions (\ref{eq:bz1}-\ref{eq:bz0}) -- is rather impractical for implementation in standard pseudo-spectral solvers. On the other hand, going back to the full primitive equations is also impractical because the latter are incompatible with periodic boundary conditions in the horizontal directions: they involve terms that are proportional to $y$ thus breaking the invariance to translations in $y$. Fortunately these terms are subdominant in the QG range of parameters and a convenient way to simulate the QG dynamics of a patch of ocean consists in discarding them from the set of Boussinesq equations. There are two such terms: first, the base state has a $z$-dependent meridional buoyancy gradient, and the vertical advection of the associated buoyancy field leads to coefficients that are proportional to $y$. We neglect this subdominant vertical advection of the background meridional buoyancy gradient. Secondly, the planetary vorticity gradient leads to a stretching term of the form $(f_0+\beta y)\partial_z {\bf u}$ in the vorticity equation. We neglect the subdominant contribution $\beta y \partial_z {\bf u}$ in the following. Specifically, we end up with the following set of dimensionless primitive-like equations for the departure fields:
\begin{eqnarray}
& & \partial_t {\bf u} + U'(z) w \, {\bf e}_x + U(z) \partial_x {\bf u} + ({\bf u}\cdot \bnabla){\bf u} + {\bf e}_z \times {\bf u} -\beta \left[\psi {\bf e}_y + \Delta_{\perp}^{-1} \{ \psi_{yz} \} {\bf e}_z \right]   \label{eq:unum} \\
\nonumber & & \qquad =  - \bnabla p + b \, {\bf e}_z + E_\perp \Delta_\perp {\bf u} + E_z \partial_{zz} {\bf u} \, , \\
& & \partial_t b + U(z) b_x - U'(z) v + N^2(z) w + {\bf u}\cdot \bnabla b = E_{b,\perp} \Delta_\perp b + E_{b,z} \partial_{zz} b \, , \label{eq:bnum}
\end{eqnarray}
where ${\bf u}=(u,v,w)$ now denotes the velocity departure from the base state. The toroidal streamfunction $\psi(x,y,z,t)$ has a vanishing horizontal area average at every depth $z$ and is defined as $\psi=\Delta_\perp^{-1} \{ \partial_y u - \partial_x v \}$. For rapid rotation and strong stratification the set of equations (\ref{eq:unum}-\ref{eq:bnum}) reduces to the limiting QG system (equation (\ref{eq:consq}) with the boundary conditions (\ref{eq:bz1}-\ref{eq:bz0})). Additionally, the form of the $\beta$ term ensures conservation of mechanical energy. 

We solve equations (\ref{eq:unum}-\ref{eq:bnum}) inside a horizontally periodic domain with the pseudo-spectral solver Coral~\citep{Miquel21coral}, previously used for the Eady model~\citep{Gallet2022} and for turbulent convective flows~\citep{Miquel20,Bouillaut21}, and validated against both analytical results~\citep{miquelPRF19} and solutions computed with the Dedalus software~\citep{Dedalus}. 
The background stratification is strong and the global rotation is fast (low Rossby number) to ensure a strongly QG regime. We use insulated boundary conditions at top and bottom for $b$, free-slip boundary conditions at the surface for the velocity departure, and a frictional boundary condition at the bottom, $\partial_z u|_{-1}=-(\tilde{\kappa}/E_z) u|_{-1}$, $\partial_z v|_{-1}=-(\tilde{\kappa}/E_z) v|_{-1}$ and $w|_{-1}=0$. Such parameterized bottom drag is detailed in the study of the Eady model~\citep{Gallet2022} together with the connection between the coefficient $\tilde{\kappa}$ and the QG friction coefficient $\kappa$ arising in equation (\ref{eq:bz0}). The dissipative coefficients in the simulation have values $\tilde{\kappa}=4.5\times 10^{-4}$, $E_z=3\times 10^{-6}$, $E_\perp=0.003$, $E_{b,z}=3 \times 10^{-7}$, $E_{b,\perp}=0.001$.








\bibliographystyle{jfm}

\bibliography{GMR}

\end{document}